\documentclass[10pt,tbtags,conference]{IEEEtran}
\IEEEoverridecommandlockouts

\usepackage[IEEE]{enderheader}
\usepackage[GMACWT,WT]{enderspecheader}
\title{The Gaussian multiple access wire-tap channel: wireless secrecy and cooperative jamming}
\author{
\authorblockN{Ender~Tekin}
\authorblockA{
Wireless Communications and Networking Laboratory\\
Electrical Engineering Department \\
The Pennsylvania State University  \\
University Park, PA 16802\\
tekin@psu.edu \vspace{-.3in}}
\and
\authorblockN{Aylin~Yener}
\authorblockA{
Wireless Communications and Networking Laboratory\\
Electrical Engineering Department \\
The Pennsylvania State University  \\
University Park, PA 16802\\
yener@ee.psu.edu
\vspace{-.3in}}
}
\begin{document}
\maketitle

\begin{abstract}
\label{sec:abstract}
We consider the General Gaussian Multiple Access Wire-Tap Channel (GGMAC-WT).  In this scenario,
multiple users communicate with an intended receiver in the presence of an intelligent and informed eavesdropper. We define two suitable secrecy measures, termed individual and collective, to reflect the confidence in the system for this multi-access environment.  We determine achievable rates such that secrecy to some pre-determined degree can be maintained, using Gaussian codebooks. We also find outer bounds for the case when the eavesdropper receives a degraded version of the intended receiver's signal. In the degraded case, Gaussian codewords are shown to achieve the sum capacity for collective constraints.  In addition, a TDMA scheme is shown to also achieve sum capacity for both sets of constraints.  Numerical results showing the new rate region are presented and compared with the capacity region of the Gaussian Multiple-Access Channel (GMAC) with no secrecy constraints.  We then find the secrecy sum-rate maximizing power allocations for the transmitters, and show that a \ital{cooperative jamming} scheme can be used to increase achievable rates in this scenario.
\end{abstract}

\section{Introduction}
\label{sec:intro}
Shannon, in \cite{shannon:secrecy}, analyzed secrecy systems in communications and showed that to achieve perfect secrecy of communications, the conditional probability of the \ital{cryptogram given a message} must be independent of the actual transmitted message. In \cite{wyner:wiretap}, Wyner applied this concept to the discrete memoryless channel, with a wire-tapper who has access to a degraded version of the intended receiver's signal.  He measured the amount of ``secrecy" using the conditional entropy $\Delta$, the conditional entropy of the transmitted message given the received signal at the wire-tapper.  In \cite{wyner:wiretap}, the region of all possible $(R,\Delta)$ pairs was determined, and the existence of a \ital{secrecy capacity}, $C_s$, for communication below which it is possible to transmit zero information to the
wire-tapper was shown.

Carleial and Hellman, in \cite{hellman-carleial:wiretap}, showed that
it is possible to send several low-rate messages, each completely
protected from the wire-tapper individually, and use the channel at close to capacity.  However, if any of the messages are available to the wire-tapper, the secrecy of the rest may also be compromised.  In \cite{leung-hellman:gaussianwiretap}, the authors extended Wyner's results in \cite{wyner:wiretap} and Carleial and Hellman's results in \cite{hellman-carleial:wiretap} to Gaussian channels.

Csisz\'ar and K\"orner, in \cite{csiszar-korner:confbroadcast},
showed that Wyner's results can be extended to weaker, so called ``less noisy" and ``more capable" channels. Furthermore, they analyzed the more general case of sending common information to both the receiver and the wire-tapper, and private information to the receiver only.

It was argued in \cite{maurer-wolf:weaktostrongsecrecy}, that the secrecy constraint developed by Wyner and later utilized by Csisz\'ar and K\"orner was ``weak" since it only constrained the rate of information leaked to the wire-tapper, rather than the total information.  It was shown that Wyner's scenario could be extended to ``strong" secrecy using extractor functions with no loss in achievable rates, where the secrecy constraint is placed on the total information obtained by the wire-tapper, as the information of interest might be in the small amount leaked.

Maurer, \cite{maurer:secretkeypublicdiscussion}, and Bennett et. al., \cite{bennettetal:genprivacy}, later focused on the process of ``distilling" a secret key between two parties in the presence of a wire-tapper utilizing a source of common randomness. In this scenario, the wire-tapper has partial information about a common random variable shared by the two parties, and the parties use their knowledge of the wire-tapper's limitations to distill a secret key.  Reference \cite{maurer:secretkeypublicdiscussion} showed that for the case when the wire-tap channel capacity is zero between two users, the existence of a ``public" feedback channel that the wire-tapper can also observe, enables the two parties to be able to generate a secret key with perfect secrecy.

In \cite{ahlswede-csiszar:CR1} and \cite{ahlswede-csiszar:CR2}, the \ital{secrecy key} capacities and \ital{common randomness} capacities, the maximum rates of common randomness that can be generated by two terminals, were developed for several models. Csisz\'ar and Narayan extended Ahslwede and Csisz\'ar's previous work to multiple-terminals by looking at what a helper terminal can contribute in \cite{csiszar-narayan:CRhelper}, and the case of multiple terminals where an arbitrary number of terminals are trying to distill a secret key and a subset of these terminals can act as helper terminals to the rest in \cite{csiszar-narayan:secrecycap-multi}. Venkatesan and Anantharam examined the cases where the two terminals generating common randomness were connected via discrete memoryless channels (DMC's) in \cite{venkatesan-anantharam:CRcap-pair}, and later generalized this to a network of DMC's connecting any finite number of terminals in \cite{venkatesan-anantharam:CRcap-network}. 
 
More recently, the notion of the wire-tap channel has been extended to parallel channels, \cite{yamamoto:secretsharing, yamamoto:secretsharinggaussian}, relay channels, \cite{oohama:relaywiretap}, and fading channels, \cite{barros:fadingwiretap}.  Fading and parallel channels were examined together in \cite{liang:Allerton06a, zang:parallelsecrecy}. Broadcast and interference channels with confidential messages were considered in \cite{liuetal:IBCconf}.  References \cite{liang:genMACconfPAP, liuetal:MACconf} examined the multiple access channel with confidential messages, where two transmitters try to keep their messages secret from each other while communicating with a common receiver. In \cite{liang:genMACconfPAP}, an achievable region is found in general, and the capacity region is found for some special cases.

In this paper, we consider the General Gaussian Multiple Access Wire-Tap Channel (GGMAC-WT), and present our results to date under the fairly general model of a wireless channel through which each user transmits open and confidential messages. We consider two separate secrecy constraints, which we call the \ital{individual} and \ital{collective} secrecy constraints, to reflect the differing amounts of confidence that users can place on the network, as defined in \cite{tekin:ASILOMAR05}.  These two different sets of security constraints are (i) the normalized entropy of any set of messages conditioned on the transmitted codewords of the other users and the received signal at the wire-tapper, and (ii) the normalized entropy of any set of messages conditioned on the wire-tapper's received signal.  Individual constraints are more conservative to ensure secrecy of any group of users even when the remaining users are compromised.  Collective constraints, on the other hand, rely on the secrecy of all users, and as such enable an increase in the achievable secrecy rates.  In \cite{tekin:ASILOMAR05}, we considered perfect secrecy for both constraints for the degraded wire-tapper case.  In \cite{tekin:ISIT06, tekin:IT06a}, we examined the achievable rates when we relaxed our secrecy constraints so that a certain amount $0 \le \delta \le 1$ of the total information was to be kept secret for the degraded case.  We also found outer bounds for the secrecy rates, and showed using collective secrecy constraints, the Gaussian codebooks achieve sum capacity.  In addition TDMA was shown to be optimal for both constraints and achieve sum-capacity.  In \cite{tekin:ALLERTON06}, we considered the General (non-degraded) GMAC and found an achievable secrecy/rate region.  In this case, we were also presented with a sum-rate maximization problem, as the maximum achievable rate depends on the transmit powers.  We noted that users may trade secrecy rates such that even ``bad" users may achieve positive secret rates at the behalf of the ``good" users.  In addition, we found the sum-rate maximizing power allocations.  We also introduced the notion of a subset of users jamming the eavesdropper to help increase the secrecy sum-rate.  This notion, which we term \ital{cooperative jamming}, is considered in detail in this paper.

\section{Main Results}
\label{sec:summary}
Our main contributions in this area are listed below,
\begin{enumerate}
\item We define two sets of information theoretic secrecy measures for a multiple-access channel:
	\begin{itemize}
	\item Individual: Secrecy is maintained for any user even if the remaining users are compromised.
	\item Collective: Secrecy is achieved with the assumption that all users are secure.
	\end{itemize}
\item Using Gaussian codebooks, we find achievable regions for both sets of constraints.  These rates may be strengthened as in \cite{maurer-wolf:weaktostrongsecrecy} to get strong secret key rates.
\item For the degraded case, we find outer bounds for both sets of constraints and show that the sum capacity bound is the same for both sets of constraints.
	\begin{itemize}
	\item For individual constraints, the achievable region is a subset of the outer bounds, but using TDMA it is possible to achieve the sum capacity.
	\item For collective constraints, it is shown that Gaussian codebooks achieve the sum capacity.
	\end{itemize}
	These outer bounds are ``strong" in the sense of \cite{maurer-wolf:weaktostrongsecrecy}, and hence we determine the strong secrecy key sum-capacities when the eavesdropper is degraded.
\item When the transmitters only have secret messages to send, we determine the power allocations that maximize the secrecy sum-rate.
\item We show that a scheme where users cooperate, with ``bad" users helping ``better" users by jamming the eavesdropper, may achieve higher secrecy rates or allow the ``better" user to achieve a positive secrecy capacity.  We term this scheme \ital{cooperative jamming}.
\end{enumerate}
\section{System Model and Problem Statement}
\newcommand{\NMt}{\tilde{\Nm}_\Mch}
\newcommand{\NWt}{\tilde{\Nm}_\Wch}
\label{sec:system}
\begin{figure}[ht]
\begin{center}
\resizebox{3in}{!}{\input{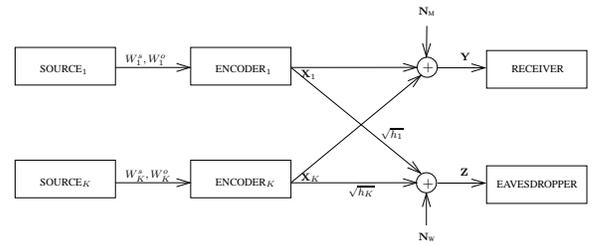}}
\caption{The standardized GMAC-WT system model}
\label{fig:gmacwt2}
\end{center}
\end{figure}
We consider $K$ users communicating with an intended receiver in the presence
of an intelligent and informed eavesdropper.  Each transmitter $k \in \Ks \triangleq \{1,2,\dotsc,K\}$ chooses a secret message $\Wsec_k$ from a set of equally likely messages $\Wssec_k=\{1, \dotsc, \Msec_k\}$, and an open message $\Wpub_k$ from a set of equally likely messages $\Wspub_k=\{1,\dotsc,\Mpub_k\}$. Let $M_k \triangleq \Msec_k \Mpub_k$, $W_k \triangleq (\Wsec_k,\Wpub_k)$, and $\Ws_k \triangleq \Wssec_k \times \Wspub_k$.  The messages are encoded into \n-length codes $\{\tilde X_k^n(W_k)\}$.  The encoded messages $\{\tilde \Xm_k\}=\{\tilde X_k^n\}$ are then transmitted, and the intended receiver and the wire-tapper each get a copy $\Ym=Y^n$ and $\Zm=Z^n$.  The receiver decodes $\Ym$ to get an estimate of the transmitted messages, $\Wmh$.  We would like to communicate with the receiver with arbitrarily low probability of error, while maintaining perfect secrecy for the secret messages given a set of secrecy constraints to be defined shortly.  By intelligent and informed eavesdropper, we mean that the channel parameters are universally known, including at the eavesdropper, and that the eavesdropper also has knowledge of the codebooks and coding scheme known.  The signals at the intended receiver and the wiretapper are given by
\begin{align}
\Ym &= \ssum_{k=1}^K \sqrt{\hM_k} \tilde \Xm_k + \NMt \\
\Zm &= \ssum_{k=1}^K \sqrt{\hW_k} \tilde \Xm_k + \NWt
\end{align}
where $\NMt,\NWt$ are the AWGN.  Each component of $\NMt \isnormal{0,\nvar_\Mch}$ and $\NWt \isnormal{0,\nvar_\Wch}$.
We also assume the following transmit power constraints:
\begin{equation}
\ninv \sumton{\tilde X_{ki}^2} \le \tilde \Pmax_k, \; k=1,\dotsc,K
\end{equation}

We examine the GGMAC-WT by an equivalent standard form, as in \cite{tekin:IT06a}:
\begin{align}
\Ym &= \ssum_{k=1}^K \Xm_k + \NM \\
\Zm &= \ssum_{k=1}^K \sqrt{h_k} \Xm_k + \NW
\end{align}
where
\begin{itemize}
\item the original codewords are scaled to get $\Xm_k = \sqrt{\frac{\hM_k}{\nvar_\Mch}}\tilde \Xm_k$.
\item The wiretapper's new channel gains are $h_k = \frac{\hW_k \nvar_\Mch}{\hM_k \nvar_\Wch}$.
\item The noises are normalized by $\NM = \frac{\NMt}{\nvar_\Mch}$ and $\NW = \frac{\NWt}{\nvar_\Wch}$.
\item The new maximum power constraints are $\Pmax_k = \frac{\hM_k}{\nvar_\Mch} \tilde \Pmax_k$.
\end{itemize}

We can show that the eavesdropper gets a stochastically degraded version of the receiver's signal if $h_1=\dotsc=h_K \equiv h < 1$.  Since the receivers do not cooperate, the capacity region depends only on the conditional marginals, and is equivalent to that of a physically degraded channel, which in turn is equivalent to $\Zm$ being a noisier version of $\Ym$:
\begin{equation}
\label{eqn:YZdeg}
\Zm=\sqrt{h} \Ym + \NMW
\end{equation}
where $\NMW \isnormal{0,(1-h)\v{I}}$.  In practical situations, we can think of this as the eavesdropper being able to wire-tap the receiver rather than receive the signals itself.

\subsection{Secrecy Measures}
\label{sec:measures}
We aim to provide each group of users with a pre-determined amount of secrecy. Letting $\Delta_\Ss$ be our secrecy constraint for any subset $\Ss$ of users, we require that $\Delta_\Ss \ge 1$ for all sets $\Ss \subseteq \Ks$.  To that end, in \cite{tekin:ASILOMAR05}, we used an approach similar to \cite{wyner:wiretap, leung-hellman:gaussianwiretap}, and defined two sets of secrecy constraints using the normalized equivocations.  These are: 
\subsubsection{Individual Constraints}
\label{sec:defind}
Define
\begin{equation}
\label{eqn:defDeltaI}
\DeltaI_k \triangleq \frac{H(\Wsec_k|\Xm_{k^c},\Zm)}{H(\Wsec_k)} \quad \forall k=1,...,K
\end{equation}
where $k^c$ is the set of all users except user $k$. If $H(\Wsec_k)=0$, we define $\DeltaI_k=1$.  $\DeltaI_k$ denotes the normalized entropy of a user's message given the received signal at the wire-tapper as well as all other users' transmitted symbols.  This constraint guarantees that information obtained at the wire-tapper about the user $k$'s signal is limited even if all other users are compromised. Let $\Wmsec_\Ss \triangleq \braces{\Wsec_k}_{k \in \Ss}$ for any set $\Ss \subseteq \Ks$ of users. Define
\begin{equation}
\label{eqn:defDeltaIS}
\DeltaI_\Ss \triangleq \frac{H(\Wmsec_\Ss|\Xm_{\Ss^c},\Zm)}{H(\Wmsec_\Ss)}
\quad \forall \Ss \subseteq \Ks=\{1,\dotsc,K\}
\end{equation}
Assume $\DeltaI_k \ge \delta$ for all users in a set $\Ss=\{1,\dotsc,S\}$. Then, we can show that
\begin{align}
H(\Wmsec_\Ss|\Xm_{\Ssc},\Zm) 
	&=\ssum_{k=1}^S H(\Wsec_k|\Wmsec_{1,\dotsc,k-1},\Xm_{\Ssc},\Zm) \\
	&\ge \ssum_{k=1}^S H(\Wsec_k|\Wmsec_{1,\dotsc,k-1},\Xm_{k^c},\Zm)\\
	&=\ssum_{k=1}^S H(\Wsec_k|\Xm_{k^c},\Zm)\\
	&\ge \ssum_{k=1}^S \delta H(\Wsec_k) \\
	&=\delta H(\Wmsec_\Ss)
\end{align}
where $\Wmsec_{1,\dotsc,k-1} \triangleq \{\Wsec_1,\dotsc,\Wsec_{k-1}\}$, and we used $\Markov{\Wsec_k}{\Xm_k}{\Zm}$.  Hence, individual constraints on each user guarantee that the constraint is satisfied for all groups of users.

\subsubsection{Collective Constraints}
\label{sec:defcol}
The individual constraints \eqref{eqn:defDeltaI} are a conservative measure as they reflect the case where users do not trust the secrecy of other users.  We next define a revised secrecy measure to take into account the multi-access nature of the channel, where there is more trust in the system, and users can count on this to achieve higher secrecy rates:
\begin{equation}
\label{eqn:defDeltaC}
\DeltaC_\Ks \triangleq \frac{H(\Wmsec_\Ks|\Zm)}{H(\Wmsec_\Ks)}
\end{equation}
which is the normalized equivocation of all the secret messages in the system.  Similar to the individual constraints case, consider this measure for an arbitrary subset $\Ss$ of users:
\begin{equation}
\label{eqn:defDeltaCS}
\DeltaC_\Ss \triangleq \frac{H(\Wmsec_\Ss|\Zm)}{H(\Wmsec_\Ss)}
\end{equation}
Assume $\DeltaC_\Ks \ge 1-\e$ for some arbitrarily small $\e$.  Then,
\begin{align}
H(\Wmsec_\Ks|\Zm) &\ge H(\Wmsec_\Ks) -\e H(\Wmsec_\Ks) \\
H(\Wmsec_\Ss|\Zm) &\ge H(\Wmsec_\Ss) + H(\Wmsec_\Ssc|\Wmsec_\Ss) -\e H(\Wmsec_\Ks) \notag \\
	&\hspace{1.2in}  - H(\Wmsec_\Ssc|\Wmsec_\Ss,\Zm) \\
	&\ge H(\Wmsec_\Ss) - \e H(\Wmsec_\Ks) \\
\DeltaC_\Ss &\ge 1-\e'
\end{align}
where $\e' \triangleq \frac{H(\Wmsec_\Ks)}{H(\Wmsec_\Ss)}\e \tozero$ as $\e \tozero$.  If $H(\Wmsec_\Ss)=0$, then we define $\Delta_\Ss=1$.  Thus, the perfect secrecy of the system implies the perfect secrecy of any group of users.  Hence, we only impose the system secrecy constraint in \eqref{eqn:defDeltaC}.  Note that in the previous section, we showed that if $\DeltaI_k \ge 1-\e$ for all $k$, then $\DeltaI_\Ks \equiv \DeltaC_\Ks \ge 1-\e$, which is why the collective constraint is strictly weaker than the individual constraint.
\subsection{Preliminary Definitions}
\label{sec:prelimdef}
\begin{definition}[Achievable rates]
\label{def:achrate}
Let $\Rm_k =(\Rsec_k,\Rpub_k)$.  The rate vector $\Rm=\paren{\Rm_1,\dotsc,\Rm_K}$ is said to be \ital{achievable} if for any given $\e>0$ there exists a code of sufficient length \n such that
\begin{align}
\ninv \log_2 \Msec_k &\ge \Rsec_k - \e \quad k=1,\dotsc,K\\
\ninv \log_2 \Mpub_k &\ge \Rpub_k - \e \quad k=1,\dotsc,K
\end{align}
and
\begin{equation}
\Perr = \frac{1}{\prod_{k=1}^K M_k} \sum_{\Wm \in {\displaystyle \times}_{k=1}^K \Ws_k}	
	\hspace{-.2in} \prob{\Wmh \neq \Wm |\Wm \text{ sent}} \le \e
\end{equation}
is the average probability of error.  In addition,
\begin{align}
\DeltaI_k &\ge 1-\e, \, \forall k \in \Ks, &&\text{if using individual constraints} \\
\DeltaC_\Ks &\ge 1-\e, &&\text{if using collective constraints}
\end{align}
We will call the set of all achievable rates $\CsI$ for individual constraints, and $\CsC$ for collective constraints.
\end{definition}
\begin{definition}[Achievable rates with $\delta$-secrecy]
\label{def:deltaachrate}
We say that $\Rmd=(\Rd_1,\dotsc,\Rd_K)$ is \ital{$\delta$-achievable} if a rate $\Rm$ is achievable such that $\Rd_k=\Rsec_k+\Rpub_k$ and $\frac{\Rsec_k}{\Rsec_k+\Rpub_k} \ge \delta$, $\forall k \in \Ks$. Since the whole message for a user, $W_k$ is uniformly distributed in $\Ws_k$, this is equivalent to stating that at least a portion $0 \le \delta \le 1$ of the message is secret for each user.  When $\delta=1$, then all users want to maintain \ital{perfect secrecy}, i.e., there is no open message.  When $\delta=0$, then the system is a standard MAC with no secret messages.
\end{definition}

Before we state our results, we also define the following:
\begin{gather}
g(\xi) \triangleq \onehalf \log_2 \paren{1+\xi}, \qquad \mmax{\xi} \triangleq \Max{\xi,0} \\
\CM_\Ss(\Pm) \triangleq g\paren{\ssum_{k\in \Ss} P_k}, \quad
\CW_\Ss(\Pm) \triangleq g\paren{\ssum_{k \in \Ss} h_k P_k} \\
\CWs_\Ss(\Pm) 
	\triangleq g\paren{\frac{\ssum_{k \in \Ss} h_k P_k}{1+\ssum_{k \in \Ssc} h_k P_k}}, \quad 
P_\Ss \triangleq \sum_{k \in \Ss} P_k\\
\Ps \triangleq \braces{\Pm=(P_1,\dotsc,P_K) \colon 0 \le P_k \le \Pmax_k,\,\forall k \in \Ks}
\end{gather}
where it should be noted that $\CM$,$\CW$ and $\CWs$ are functions of the transmit powers, even when it is not made explicit in the text to simplify notation.
\section{Achievable Secrecy Rate Regions}
\label{sec:ach}
\newcommand{\GsT}{\Gs[T]}
\newcommand{\GsTd}{\GsT_\delta}
In this section, we find a set of achievable rates using Gaussian codebooks and simultaneous superposition coding as described in Appendix \ref{app:achprfenc}, which we call \GsI for individual constraints, and \GsC for collective constraints.  We also find a region achievable using TDMA, and is valid for both sets of constraints.  This region, which we call \GsT, is a subset of the achievable region when using collective constraints, but enlarges the achievable region when using individual constraints.  We should also note that these rates can be strengthened using extractor functions as shown in \cite{maurer-wolf:weaktostrongsecrecy}, for details see \cite{tekin:IT07a}.

\subsection{Individual Secrecy}
In \cite{leung-hellman:gaussianwiretap}, it has been shown that
Gaussian codebooks can be used to maintain secrecy for
a single user wire-tap channel. Using a similar approach, we show
that an achievable region using individual constraints is given by:
\begin{theorem}
\label{thm:achind}
Define $\GsI(\Pm) =$
\begin{equation}
\label{eqn:achindP}
\braces{\Rm \colon \begin{cases}
\sum_{k \in \Ss} \Rsec_k \le \mmax{\CM_\Ss-\sum_{k\in\Ss}\CW_k}, &\forall \Ss \\
\sum_{k \in \Ss} \paren{\Rsec_k+\Rpub_k} \le \CM_\Ss, & \forall \Ss 
\end{cases}}
\end{equation}
Then, the region 
\begin{equation}
\label{eqn:achind}
\GsI = \text{convex closure of} \bigcup_{\Pm \in \Ps} \GsI(\Pm)
\end{equation}
is achievable with individual constraints.
\end{theorem}
\begin{corollary}
\label{cor:achindd}
Define $\GsdI(\Pm) =$
\begin{equation}
\label{eqn:achinddP}
\braces{\Rmd \colon \begin{cases}
\sum_{k \in \Ss} \Rd_k \le \deltainv \mmax{\CM_\Ss-\sum_{k\in\Ss}\CW_k}, &\forall \Ss \\ 
\sum_{k \in \Ss} \Rd_k \le \CM_\Ss, & \forall \Ss 
\end{cases}}
\end{equation}
Then, the region
\begin{equation}
\label{eqn:achindd}
\GsdI = \text{convex closure of} \bigcup_{\Pm \in \Ps} \GsdI(\Pm)
\end{equation}
is $\delta$-achievable with individual constraints.
\end{corollary}
\begin{proof}
See Appendix \ref{app:achprfind}.
\end{proof}

\subsection{Collective Secrecy}
\label{sec:achcol}
In this section, we give an achievable rate region for collective constraints.  Our main result is:
\begin{theorem}
\label{thm:achcol}
Define $\GsC(\Pm) = $
\begin{equation}
\label{eqn:achcolP}
\braces{\Rm \colon \begin{cases}
\sum_{k=1}^K \Rsec_k \le \mmax{\CM_\Ks-\CW_\Ks} &\\
\sum_{k \in \Ss} \paren{\Rsec_k+\Rpub_k} \le \CM_\Ss, & \forall \Ss
\end{cases}}
\end{equation}
Then, the region 
\begin{equation}
\label{eqn:achcol}
\GsC = \text{convex closure of} \bigcup_{\Pm \in \Ps} \GsC(\Pm)
\end{equation}
is achievable with collective constraints.
\end{theorem}
\begin{corollary}
\label{cor:achcold}
Define $\GsdC(\Pm) = $
\begin{equation}
\label{eqn:achcoldP}
\braces{\Rmd \colon \begin{cases}
\sum_{k=1}^K \Rd_k \le \deltainv \mmax{\CM_\Ks-\CW_\Ks}, &\forall k \\
\sum_{k \in \Ss} \Rd_k \le \CM_\Ss, & \forall \Ss
\end{cases}}
\end{equation}
Then, the region
\begin{equation}
\label{eqn:achcold}
\GsdC = \text{convex closure of} \bigcup_{\Pm \in \Ps} \GsdC(\Pm)
\end{equation}
is $\delta$-achievable with collective constraints.
\end{corollary}
\begin{proof}
See Appendix \ref{app:achprfcol}.
\end{proof}

\subsection{Time-Division Multiple-Access}
\label{sec:achTDMA}
We can also use TDMA to get an achievable region.  Consider this scheme: Let $\alpha_k \in [0,1], \, k=1,\dotsc,K$ and $\sum_{k=1}^K \alpha_k = 1$.  User $k$ only transmits $\alpha_k$ of the time with power $P_k/\alpha_k$, hence satisfying the average power constraints. The transmission uses the scheme described in \cite{leung-hellman:gaussianwiretap}.  Since only one user is transmitting at a given time, both sets of constraints collapse down to a set of single-user secrecy constraints, for which the results were given in \cite{leung-hellman:gaussianwiretap}:
\begin{theorem}
\label{thm:achTDMA}
Define,
\begin{equation}
\label{eqn:achTDMAP}
\GsT(\Pm,\vg{\alpha}) \triangleq \braces{\Rm \colon \hspace{-.05in} \begin{cases}
\Rsec_k \le \alpha_k \mmax{g\paren{\frac{(1-h_k)P_k}{\alpha_k + h_k P_k}}}, 
	& \hspace{-.05in}\forall k\\
\Rsec_k+\Rpub_k \le \alpha_k g\paren{\frac{P_k}{\alpha_k}}, & \hspace{-.05in} \forall k 
\end{cases}}
\end{equation}
Then, the following set of rates is achievable:
\begin{equation}
\label{eqn:achTDMA}
\GsT=\text{convex closure of} \bigcup_{\Pm \in \Ps}
\bigcup_{\substack{\v{0} \preceq \vg{\alpha} \preceq \v{1} \\ \sum_{k=1}^K \alpha_k=1}}
\GsT(\Pm,\vg{\alpha})
\end{equation}
\end{theorem}
\begin{corollary}
\label{cor:achTDMAd}
Define
\begin{equation}
\label{eqn:achTDMAdP}
\hspace{-.00in}
\GsTd(\Pm,\vg{\alpha}) \triangleq \braces{\Rmd \colon \hspace{-.05in} \begin{cases}
\Rd_k \le \frac{\alpha_k}{\delta} \mmax{g\paren{\frac{(1-h_k)P_k}{\alpha_k + h_k P_k}}}, 
	& \hspace{-.08in} \forall k\\
\Rd_k \le \alpha_k g\paren{\frac{P_k}{\alpha_k}}, & \hspace{-.08in} \forall k 
\end{cases}} \hspace{-.03in}
\end{equation}
Then, the region
\begin{equation}
\label{eqn:achTDMAd}
\GsTd=\text{convex closure of} \bigcup_{\Pm \in \Ps}
\bigcup_{\substack{\v{0} \preceq \vg{\alpha} \preceq \v{1} \\ \sum_{k=1}^K \alpha_k=1}}
\GsTd(\Pm,\vg{\alpha})
\end{equation}
is $\delta$-achievable.
\end{corollary}
\begin{proof}
See Appendix \ref{app:achprfTDMA}.
\end{proof}
For collective secrecy constraints, \GsT is a subset of \GsC.  For individual secrecy constraints, however, this region is sometimes a superset of \GsI, and sometimes a subset of \GsI, but most of the time it helps enlarge this region.  We can then, using time-sharing arguments, find a new achievable region for individual constraints that is the convex-closure of the union of the two regions, i.e.,
\begin{proposition}
The following region is achievable for individual secrecy constraints:
\begin{equation}
\label{prop:achI}
\GsI_\cup = \text{convex closure of }\paren{\GsI \cup \GsT}
\end{equation}
\end{proposition}

\newcommand{\CsCu}{\Csu[C]}
\newcommand{\GsCu}{\Gsu[C]}
\newcommand{\CsIu}{\Csu[I]}
\newcommand{\GsIu}{\Gsu[I]}
\newcommand{\CsdCu}{\Csu[C]_\delta}
\newcommand{\GsdCu}{\Gsu[C]_\delta}
\newcommand{\CsdIu}{\Csu[I]_\delta}
\newcommand{\GsdIu}{\Gsu[I]_\delta}
\section{Outer Bounds For Degraded Eavesdropper}
\label{sec:out}
In this section, we present outer bounds on the sets of achievable secrecy rates \ital{for the degraded case}.  We find the \ital{secrecy sum-capacity} which is equal for both sets of constraints, and show the region given in Theorem \ref{thm:achcol} achieves this capacity, as does the TDMA region given in Theorem \ref{thm:achTDMA}.

\subsection{Individual Secrecy}
\begin{theorem}
\label{thm:outind}
For the GMAC-WT, given the set of transmit powers $\Pm$, the achievable rates for individual constraints belong to the region
\begin{equation}
\label{eqn:outindP}
\GsIu(\Pm) = \braces{\Rm \colon \begin{cases}
\Rsec_k \le \CM_k-\CW_k, &\forall k \\
\sum_{k \in \Ss} \paren{\Rsec_k+\Rpub_k} \le \CM_\Ss, & \forall \Ss 
\end{cases}}
\end{equation}
\end{theorem}
\begin{corollary}
\label{cor:outindd}
The $\delta$-achievable rates must be in the region below:
\begin{equation}
\label{eqn:outinddP}
\GsdIu(\Pm) = \braces{\Rmd \colon \begin{cases}
\Rd_k \le \deltainv \paren{\CM_k-\CW_k}, &\forall k\\
\sum_{k \in \Ss} \Rd_k \le \CM_\Ss, & \forall \Ss 
\end{cases}}
\end{equation}
\end{corollary}
\begin{proof}
See Appendix \ref{app:outprfind}.
\end{proof}
\subsection{Collective Secrecy}
Our main result is presented in the following theorem:
\begin{theorem}
\label{thm:outcol}
For the GMAC-WT with collective secrecy constraints, given the transmit power $\Pm$, the secure rate-tuples must be in the region
\begin{equation}
\label{eqn:outcolP}
\GsCu(\Pm) = \braces{\Rm \colon \begin{cases}
\sum_{k=1}^K \Rsec_k \le \CM_\Ks-\CW_\Ks &\\
\sum_{k \in \Ss} \paren{\Rsec_k+\Rpub_k} \le \CM_\Ss, & \forall \Ss 
\end{cases}}
\end{equation}
\end{theorem}
\begin{corollary}
\label{cor:outcold}
The $\delta$-achievable rates must be in the region
\begin{equation}
\label{eqn:outcoldP}
\GsdCu(\Pm) = \braces{\Rmd \colon \hspace{-.05in}\begin{cases}
\sum_{k=1}^K \Rd_k \le \deltainv \paren{\CM_\Ks-\CW_\Ks}, &\forall k \\
\sum_{k \in \Ss} \Rd_k \le \CM_\Ss, & \forall \Ss 
\end{cases}} \hspace{-.01in}
\end{equation}
\end{corollary}
\begin{proof}
See Appendix \ref{app:outprfcol}.
\end{proof}

\subsection{Secrecy Sum-Capacity}
For the degraded case, we can find the secrecy sum-capacities for both sets of constraints.  Incidentally, the secrecy sum-capacity is the same, and is stated below:
\begin{theorem}
For the degraded case, the secrecy sum capacity is given by
\begin{equation}
\sum_{k=1}^K \Rsec_k \le \CM_\Ks-\CW_\Ks = g\paren{\frac{(1-h)P_\Ks}{1+hP_\Ks}}
\end{equation}
\end{theorem}
\begin{proof}
See Appendix \ref{app:degsumcap}
\end{proof}
The converses proven in this section for the degraded case are strong converses in the sense of \cite{maurer-wolf:weaktostrongsecrecy}.  Strengthening the achievable rates as shown in the same paper thus establishes the \ital{strong secret key sum-capacity} for the degraded case.
\newcommand{\Pmopt}{\Pm^*}
\section{Maximization of Sum Rate for Collective Constraints \& Cooperative Jamming}
\label{sec:summax}
Clearly, the collective secrecy constraints are more interesting in the sense that they utilize the multi-access nature of the channel.  When we impose individual constraints, each user has to fend for itself, confusing the eavesdropper without depending on the other users.  However, this only allows ``good" users to be able to communicate.  Collective constraints, on the other hand, allow users to help each other, and achieve a larger rate region.  Thus, in this section we concentrate on collective constraints.

The achievable region given in Theorem \ref{thm:achcol} depends on the transmit powers.  We are naturally interested in the power allocation $\Pmopt=(\Popt_1,\dotsc,\Popt_\Ks)$ that would maximize the total secrecy sum-rate.  For ease of illustration, we consider the $K=2$ user case, and assume $h_1 \le h_2$.  In other words, user $1$'s channel is ``better" than user $2$'s channel since a lower channel gain means less information leaks to the eavesdropper.
\subsection{Sum-Rate Maximization}
\label{sec:summaxopt}
We would like to find the power allocation that will maximize the secrecy sum-rate achievable found in Theorem \ref{thm:achcol}.  Stated formally, we are interested in
\begin{align}
\hspace{-.05in}
\max_{\Pm \in \Ps} \; \CM_\Ks -\CW_\Ks 
	&= \max_{\Pm \in \Ps} \; g\paren{P_1+P_2} - g\paren{h_1 P_1+h_2 P_2} \hspace{-.05in} \\
	&\equiv \min_{\Pm \in \Ps} \; \rho(\Pm) \label{eqn:sumprob1}
\end{align}
where $\rho(\Pm)\triangleq \frac{1+h_1 P_1+h_2 P_2}{1+P_1+P_2}$, and we used the monotonicity of the $\log$ function.  The solution to this problem is given below:
\begin{theorem}
The secrecy sum-rate maximizing powers are
\label{thm:summaxopt}
\begin{equation}
(\Popt_1,\Popt_2) = 
	\begin{cases}
	(\Pmax_1,\Pmax_2), & \text{if } h_1 < 1, \, h_2 < \frac{1+h_1 \Pmax_1}{1+\Pmax_1} \\
	(\Pmax_1,0), & \text{if } h_1 < 1, \, h_2 \ge \frac{1+h_1 \Pmax_1}{1+\Pmax_1} \\
	(0,0), & \text{otherwise}
	\end{cases}
\end{equation}\end{theorem}
\begin{proof}
See Appendix \ref{app:summaxprfopt}.
\end{proof}
This result is easily generalized to $K>2$ users, see \cite{tekin:IT07a}.

\newlength{\figsize}
\setlength{\figsize}{2in}
\begin{figure}[t]
\centering
\subfigure[individual constraints]{\label{fig:Regdelta-h05I} 	
	\includegraphics[width=\figsize,angle=0]{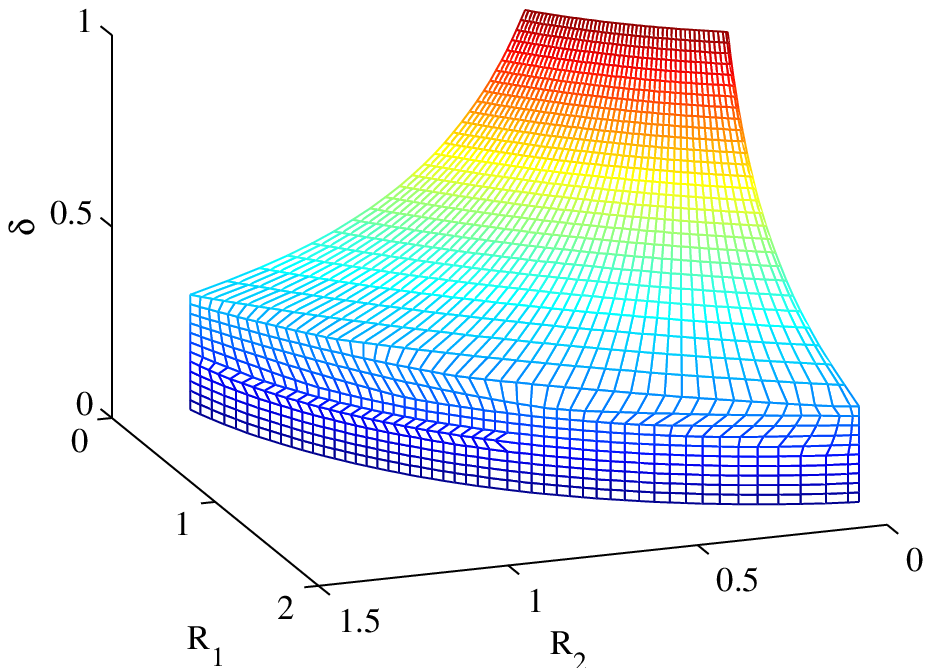}}
\subfigure[collective constraints]{\label{fig:Regdelta-h05C} 
	\includegraphics[width=\figsize,angle=0]{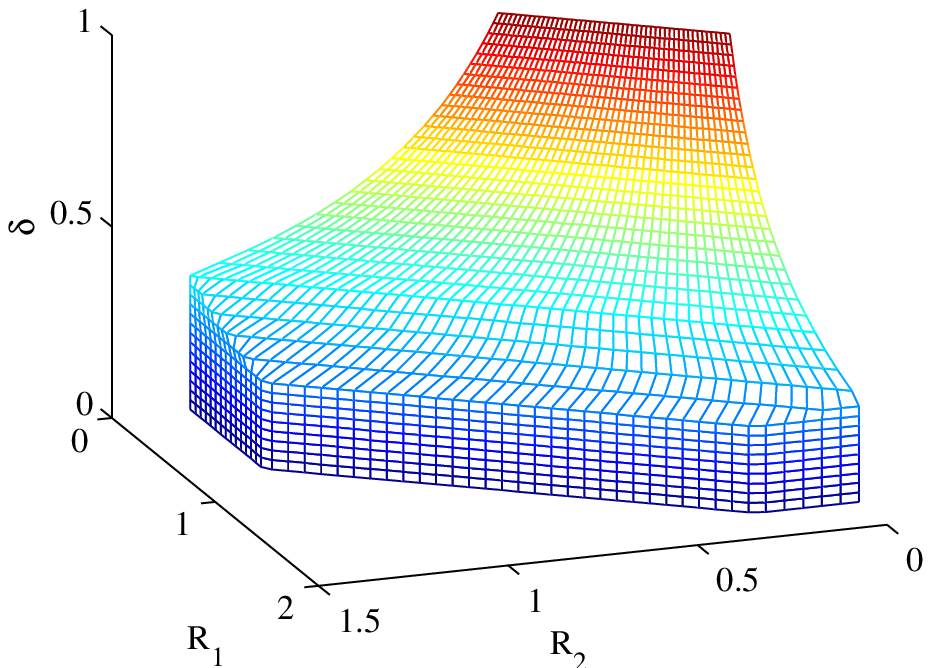}}
\vspace{-.1in}
\caption{The two-user rate region vs $\delta$.  $h=0.5$.}
\label{fig:Regdelta}
\vspace{-.3in}
\end{figure}
\subsection{Cooperative Jamming}
\label{sec:summaxcj}
\vspace{-.05in}
The solution to the optimization problem given in Theorem \ref{thm:summaxopt} shows that when $h_2 \ge \frac{1+h_1 \Pmax_1}{1+\Pmax_1}$, which implies $g\paren{\frac{h_2 P_2}{1+h_1 \Pmax_1}} \ge g\paren{\frac{P_2}{1+\Pmax_1}}$ for all $P_2>0$, then user $2$ should not transmit as it cannot achieve secrecy.  However, such a user $k$ has high eavesdropper channel gain, $h_k$, and if it started ``jamming" the channel, then it would harm the eavesdropper more than it would the intended receiver.  Since the secrecy capacity for the remaining single user is the difference of the channel capacities, it might be possible to increase user $1$'s capacity, or even, when $h_1>1$ allow it to start transmitting.  The jamming is done simply by transmitting white Gaussian noise, i.e., $\Xm_2 \isnormal{0,P_2 \v{I}}$.  As shown in \cite{tekin:IT07a}, it is always better for ``bad" users to jam.  The problem is finding the power allocations that will maximize the secrecy capacity for user $1$, formally stated as:
\begin{equation}
\max_{\Pm \in \Ps} \quad
		g\paren{\frac{P_1}{1+P_2}}-g\paren{\frac{h_1 P_1}{1+h_2 P_2}} 
	\equiv \min_{\Pm \in \Ps} \frac{\rho(\Pm)}{\phi_2 (P_2)}
\end{equation}
where $\phi_j(P) \triangleq \frac{1+h_j P}{1+P}$.
Note that we must have $\phi_2(P_2)>1$ to have an advantage over not jamming.  In general, this scheme can be shown to achieve the following secrecy capacity:
\begin{theorem}
\label{thm:summaxcj}
The secrecy capacity using cooperative jamming is $g\paren{\frac{(1-h_1)\Popt_1+(1-h_2)\Popt_2}{1+h_1\Popt_1+h_2\Popt_2}}$ where the optimum power allocations are given by $(\Popt_1,\Popt_2)=$
\begin{equation}
\begin{cases}
	(\Pmax_1,0), &\text{if } h_1 \le 1,\, \frac{1+h_1 \Pmax_1}{1+\Pmax_1} \le h_2 \le 1 \\
	(\Pmax_1,\mmax{\min \braces{p,\Pmax_2}}), &\text{if } h_1 \le 1,\, h_2 > 1\\
	(\Pmax_1,\min \braces{p,\Pmax_2}), &\text{if } h_1 \ge 1,\, \frac{h_1-1}{h_2-h_1} < \Pmax_2 \\
	(0,0), &\text{if } h_1 \ge 1,\, \frac{h_1-1}{h_2-h_1} \ge \Pmax_2
\end{cases} \hspace{-.2in}
\end{equation}
where $p=\frac{h_1-1}{h_2-h_1} + \frac{\sqrt{h_1 h_2 (h_2-1) \bracket{(h_2-1)+(h_2-h_1)\Pmax_1}}}{h_2 (h_2-h_1)}$.
\end{theorem}
\begin{proof}
See Appendix \ref{app:summaxprfcj}.
\end{proof}
In the case unaccounted for above, when $h_1 \le 1$ and $h_2 \le \frac{1+h_1\Pmax}{1+\Pmax}$, both users should be transmitting as shown in Theorem \ref{thm:summaxopt}.  The solution shows that the jamming user should jam if it is not single-user decodable, and if it has enough power to make the other user ``good" in the new standardized channel.  For the case with $K>2$ users, see \cite{tekin:IT07a}.
\vspace{-.04in}
\section{Numerical Results}
\label{sec:results}
\vspace{-.04in}

\begin{figure}[!t]
\centering
\subfigure[individual constraints]{\label{fig:Regh-d05I}
	\includegraphics[width=\figsize,angle=0]{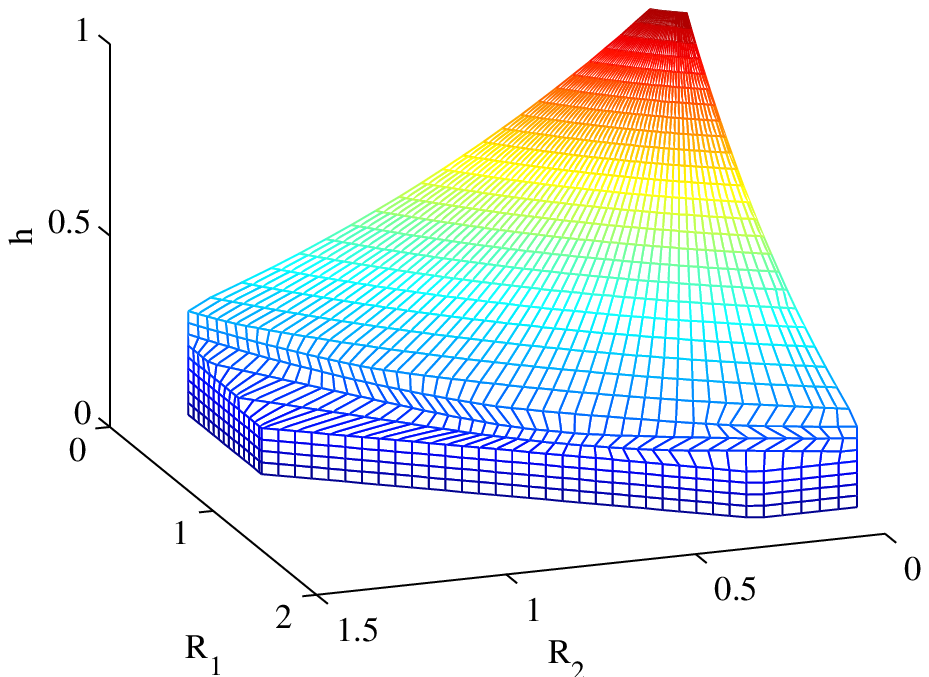}}
\subfigure[collective constraints]{\label{fig:Regh-d05C}
	\includegraphics[width=\figsize,angle=0]{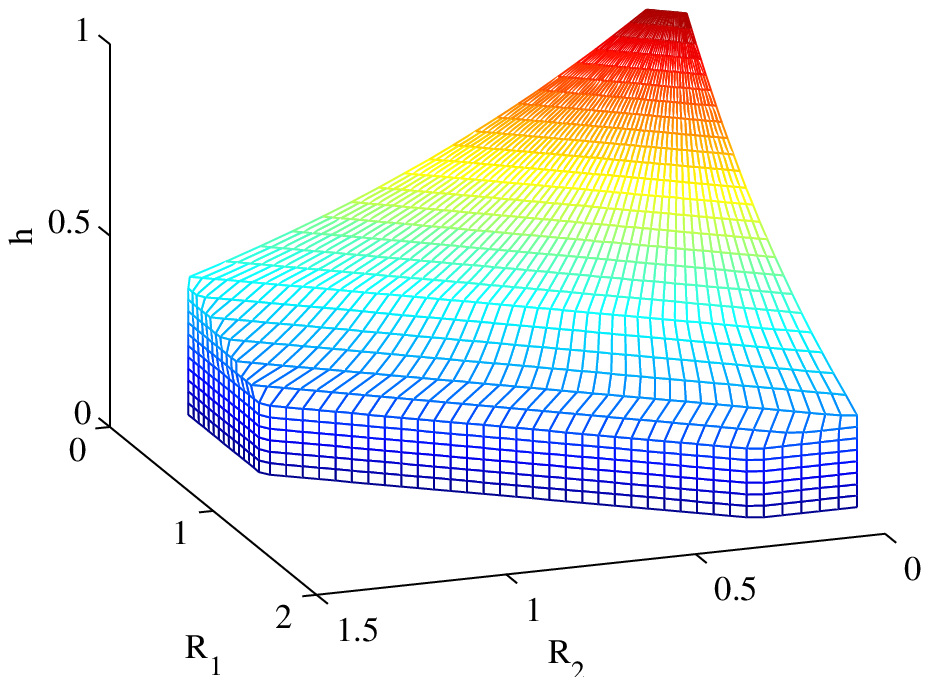}}
\vspace{-.1in}
\caption{The two-user rate region vs $h$.  $\delta=0.5$.}
\label{fig:Regh}
\vspace{-.25in}
\end{figure}

In this section, we present numerical results to illustrate the achievable rates and our cooperative jamming scheme.  To see how the channel parameters and the required level of secrecy affect the achievable rates, we consider the two-user degraded case as illustrated in Figures \ref{fig:Regdelta},\ref{fig:Regh}.  We observe that if the wire-tapper's degradedness is severe ($h \tozero$), then the secrecy sum-capacity goes to $g(P_\Ks)$, i.e., we incur no loss in sum capacity and can still communicate with perfect secrecy as the sum capacity is achievable for both sets of constraints.  On the other hand, if the wire-tapper is not severely degraded, ($h \toone$), then the secrecy sum-capacity becomes zero. Another point to note is that the $\delta$-achievable sum-secrecy capacity is limited by$\frac{1}{2\delta}\log \paren{\frac{1+P_\Ks}{1+hP_\Ks}}$, and this term is an increasing function of $P_\Ks$. However, as $P_\Ks \toinf$, it is upper bounded by $-\frac{1}{2\delta} \log h$.  We see that regardless of the available power, the sum capacity with a non-zero level of secrecy is limited by the degradedness, $h$, and the level of secrecy required, $\delta$.  

\begin{figure}[t]
\centering
\subfigure[User Powers]{\label{fig:cj1}
	\includegraphics[width=1\figsize,height=\figsize, angle=0]{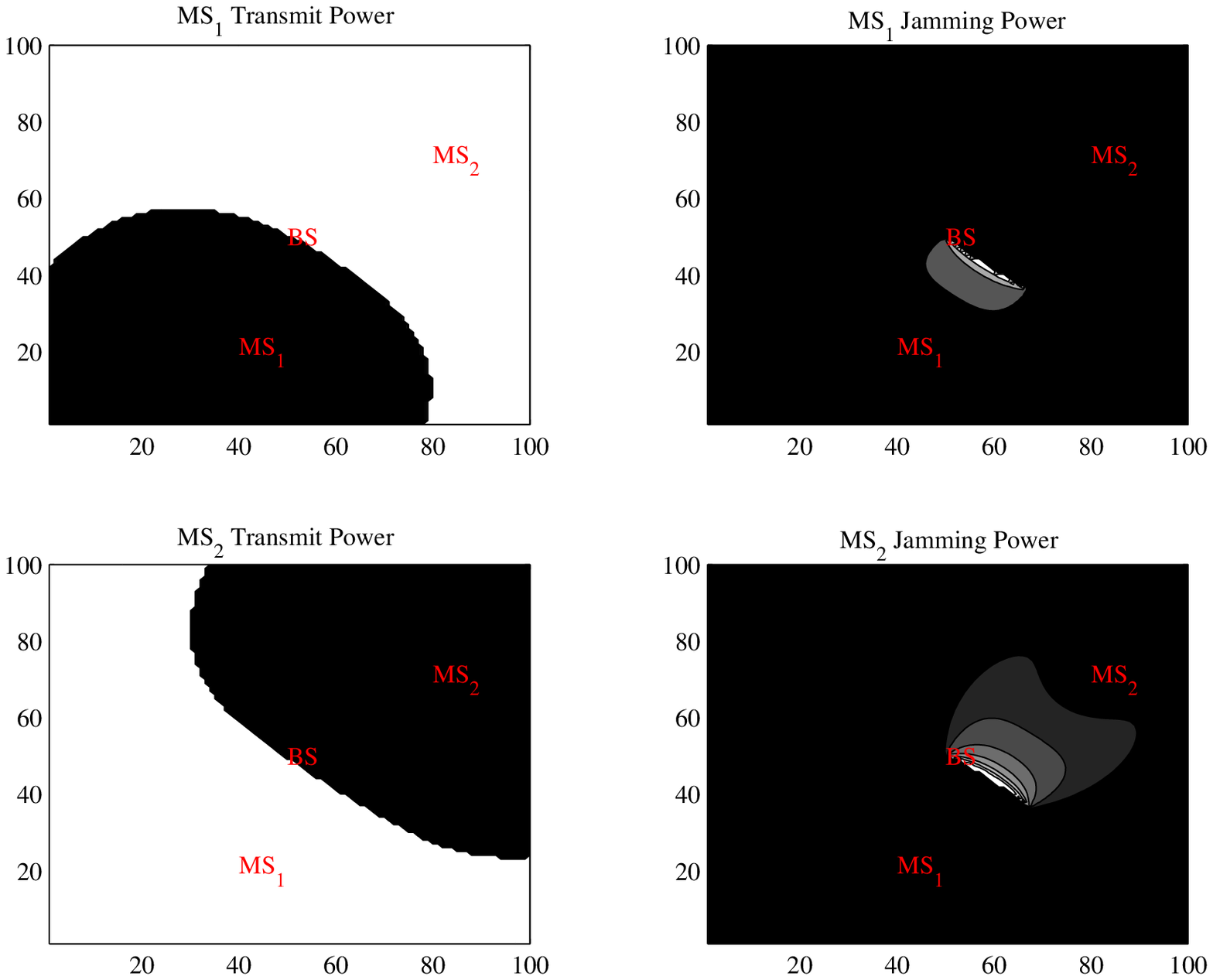}}
\subfigure[Sum-Rate]{\label{fig:cj2}
	\includegraphics[width=\figsize,height=.5\figsize,angle=0]{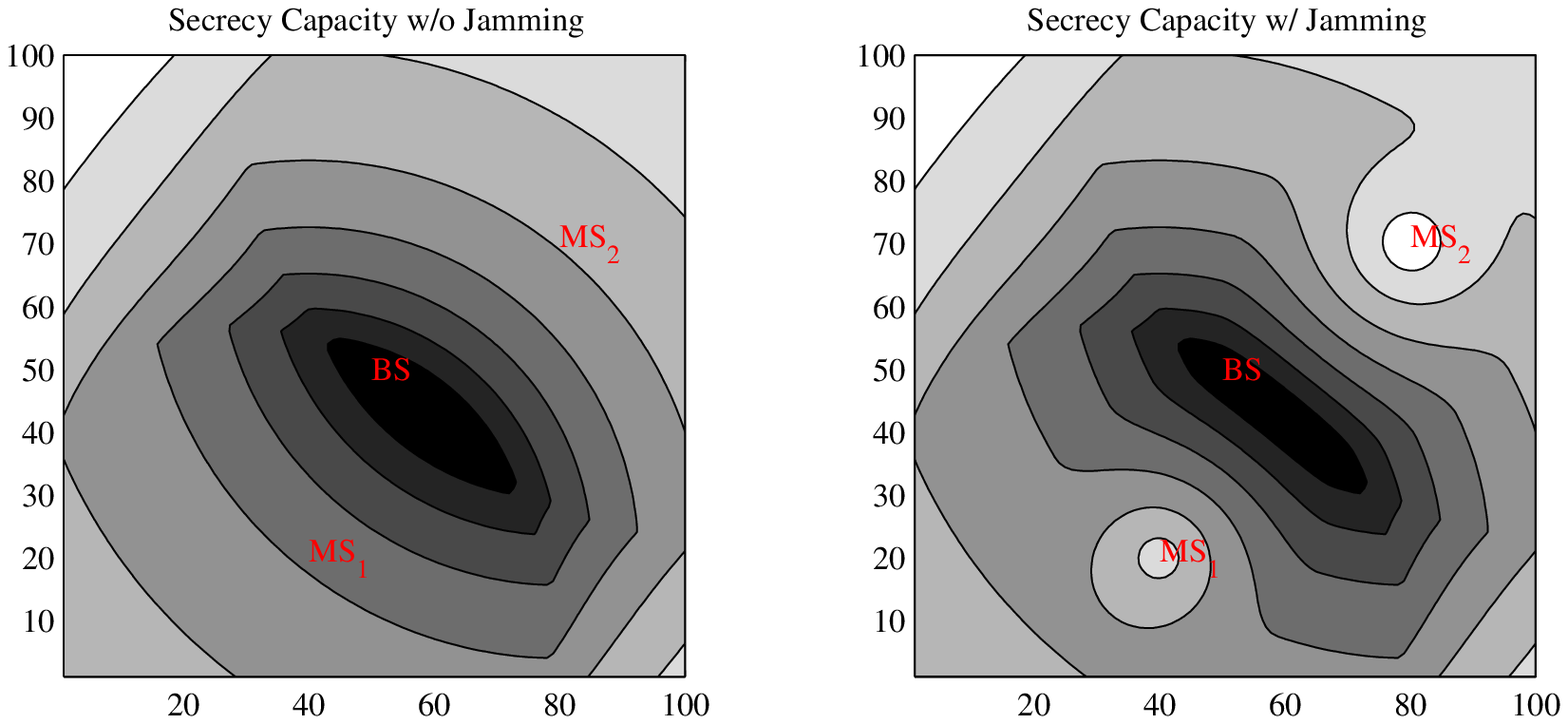} \vspace{-.1in} }
\vspace{-.1in}
\caption{\small Cooperative jamming example.}
\label{fig:CJ}
\vspace{-.2in}
\end{figure}

We also show the results of a scenario with a mobile eavesdropper (in general non-degraded) and a static base station in a $100\times100$ grid.  We use a simple path loss model, and show the optimum transmit/jamming powers when the eavesdropper is at $(x,y)$ in Figure \ref{fig:cj1}, and the resulting sum-rates achieved with and without cooperative jamming in Figure \ref{fig:cj2}, where lighter shades correspond to higher values.  Users need higher jamming powers when the eavesdropper is closer to the base station, but higher rates are achieved with less power when the eavesdropper is closer to the jammer.
Also, the area near the BS where secrecy sum-rate is zero without cooperative jamming is reduced.
\vspace{-.04in}
\section{Conclusion}
\label{sec:conclusion}
\vspace{-.04in}
In this paper, we considered the GMAC in the presence of an external eavesdropper from which information is to be kept secret.  We have established achievable rates, and outer bounds on secrecy capacity for certain scenarios.  We have shown that the multiple-access nature of the channel can be utilized to improve the secrecy of the system.  Allowing confidence in the secrecy of all users, the secrecy rate of a user is improved since the undecoded messages of any set of users acts as additional noise at the wire-tapper and precludes it from decoding the remaining set of users. We have also found the sum-rate maximizing power allocations, and show a novel scheme, which we call \ital{cooperative jamming}, which can be utilized to increase the achievable sum-rate.  We note the cooperative achievements that are possible for the GGMAC-WT with collective secrecy: (i) ``good" users may sacrifice their rates so that some ``bad" users can achieve positive secrecy rates, and (ii) really ``bad" users may sacrifice power to help the actual transmitters by jamming the eavesdropper.

\appendices
\newcommand{\Xc}{\mathfrak{X}}
\newcommand{\Xcsec}{\secsign{\Xc}}
\newcommand{\Xcpub}{\pubsign{\Xc}}
\newcommand{\Xcx}{\xsign{\Xc}}
\newcommand{\Wmi}[1][s]{\Wm^{{\scriptscriptstyle ({\scriptstyle #1})}}}
\newcommand{\XmS}{\Xm_\Sigma}
\newcommand{\Deltai}[1][s]{\Delta^{{\scriptscriptstyle ({\scriptstyle #1})}}}

\vspace{-.03in}
\section{Achievable Rates}
\label{app:achprf}
\vspace{-.03in}

\subsection{Superposition Encoding Scheme}
\label{app:achprfenc}
\vspace{-.02in}
For each user $k \in \Ks$, consider the scheme:
\begin{IEEEenumerate}[
\setlength{\topsep}{0.0in}
\setlength{\parskip}{0in}
\setlength{\labelindent}{0.06in}
\setlength{\labelwidth}{0.05in}
\setlength{\labelsep}{3pt}]

\item	Generate $3$ codebooks $\Xcsec_k,\Xcpub_k$ and $\Xcx_k$.  $\Xcsec_k$ consists of $\Msec_k$	codewords, each component of which is drawn $\isnormal{0,\secsign{\lambda}_k P_k -\varepsilon}$. Codebook $\Xcpub_k$ has $\Mpub_k$ codewords with each component randomly drawn $\isnormal{0,\pubsign{\lambda}_k P_k-\varepsilon}$ and $\Xcx_k$ has $\Mx_k$ codewords with each component randomly drawn $\isnormal{0,\xsign{\lambda}_k P_k-\varepsilon}$ where $\varepsilon$ is an arbitrarily small number to ensure that the power constraints on the codewords are satisfied with high probability and $\secsign{\lambda}_k+\pubsign{\lambda}_k +\xsign{\lambda}_k=1$. Define $\Rx_k=\ninv \log \Mx_k$, $\Mt_k=\Msec_k \Mpub_k \Mx_k$ and $\Rt_k=\ninv \log \Mt_k = \Rsec_k+\Rpub_k+\Rx_k$.

\item To transmit message $\Wm_k = (\Wsec_k,\Wpub_k) \in \Wssec_k \times \Wspub_k$, user $k$ finds the $2$ codewords corresponding to components of $\Wm_k$ and also uniformly chooses a codeword from $\Xcx_k$. He then adds all these codewords and transmits the resulting codeword, $\Xm_k$, so that we are actually transmitting one of $\Mt_k$ codewords.  Since  codewords are chosen uniformly, for each message $\Wsec_k$, we transmit one of $\Mpub_k \Mx_k$ codewords.
\end{IEEEenumerate}

\subsection{Individual Constraints}
\label{app:achprfind}
Let $\Pm \in \Ps$ and $\Rm$ satisfy \eqref{eqn:achindP}.  We choose $\{\Rx_k\}$ to satisfy:
\begin{align}
\label{eqn:achindx} \Rpub_k +\Rx_k &= \CW_k, && \forall k \in \Ks \\
\label{eqn:achindM} \ssum_{k \in \Ss} \paren{\Rsec_k+\Rpub_k + \Rx_k} &\le \CM_\Ss,
	&& \forall \Ss \subseteq \Ks
\end{align}
if $\Rsec_k>0$, and if $\Rsec_k=0$, then we set $\Rx_k=0$ and we do not impose the condition given in \eqref{eqn:achindx}.  If $\Rm$ satisfies \eqref{eqn:achindP}, we can always choose $\{\Rx\}$ to satisfy the above.

Consider the subcode $\{\Xcsec_k\}_{k=1}^K$.  From this point of view, the coding scheme described is equivalent to each user $k \in \Ks$ selecting one of $\Msec_k$ messages, and sending a uniformly chosen codeword from among $\Mpub_k \Mx_k$ codewords for each. We can thus write the following:
\begin{align}
H(\Wsec_k|\Xm_{k^c},\Zm) \hspace{-1in}&  \notag \\
  &=H(\Wsec_k,\Xm_k,\Xm_{k^c},\Zm)-H(\Xm_k|\Wsec_k,\Xm_{k^c},\Zm) \notag \\
  	&\hspace{.2in} -H(\Xm_{k^c},\Zm) \\
  &=H(\Zm|\Wsec_k,\Xm_k,\Xm_{k^c}) + H(\Xm_k,\Xm_{k^c}|\Wsec_k)+H(\Wsec_k) \notag \\
  	&\hspace{.2in} -H(\Xm_k|\Wsec_k,\Xm_{k^c},\Zm)-H(\Xm_{k^c},\Zm) \\
  &=H(\Zm|\Xm_k,\Xm_{k^c}) + H(\Xm_k,\Xm_{k^c}|\Wsec_k)+H(\Wsec_k) \notag \\
		&\hspace{.2in} -H(\Xm_k|\Wsec_k,\Xm_{k^c},\Zm)-H(\Zm|\Xm_{k^c}) -H(\Xm_{k^c}) \hspace{-.05in} \\
  &=H(\Wsec_k)+H(\Zm|\Xm_k,\Xm_{k^c})+H(\Xm_{k^c}|\Wsec_k)-H(\Xm_{k^c}) \hspace{-.1in} \\
		&\hspace{.2in} +H(\Xm_k|\Xm_{k^c},\Wsec_k)-H(\Xm_k|\Wsec_k,\Xm_{k^c},\Zm) -H(\Zm|\Xm_{k^c}) \notag \\
  &=H(\Wsec_k) - I(\Xm_k;\Zm|\Xm_{k^c}) + I(\Xm_k;\Zm|\Wsec_k,\Xm_{k^c})
  	\label{eqn:achprfind1}
\end{align}
where we used the fact that $H(\Xm_{k^c}|\Wsec_k) = H(\Xm_{k^c})$.  
By the GMAC coding theorem, we have $I(\Xm_k;\Zm|\Xm_{k^c}) \le n\CW_k$. We can also write
$I(\Xm_k;\Zm|\Wsec_k,\Xm_{k^c})= H(\Xm_k|\Wsec_k,\Xm_{k^c})- H(\Xm_k|\Wsec_k,\Xm_{k^c},\Zm)$.
The coding scheme implies that $H(\Xm_k|\Wsec_k,\Xm_{k^c}) = H(\Xm_k|\Wsec_k) = n\CW_k$.  Also, $H(\Xm_k|\Wsec_k,\Xm_{k^c},\Zm) \le n\delta_n$, where $\delta_n \tozero$ due to Fano's inequality; given $\Wsec_k$, the subcode for user $k$
is, with high probability, a ``good" code for the wiretapper.  Combining these in \eqref{eqn:achprfind1}, we can write
\begin{equation}
\label{eqn:achprfind2}
\DeltaI_k \ge 1-\frac{n \CW_k - n \CW_k + n\delta_n}{H(\Wsec_k)} =1-\e 
\end{equation}
where $\e = \frac{\delta_n}{\Rsec_k} \tozero$ as $n \toinf$.

The corollary follows simply by using the definition of $\delta$-achievability, and noting that if $\Rsec_k$ is achievable, then $\Rsec_k \ge \delta \Rd_k$ and substituting this into \eqref{eqn:achindP}.
\hfill \QED

\subsection{Collective Constraints}
\label{app:achprfcol}
The proof is similar to the proof for individual constraints.  Let $\Pm \in \Ps$ and $\Rm$ satisfy \eqref{eqn:achcolP} and assume the coding scheme is as given in Appendix \ref{app:achprfenc}.
We choose the rates such that
\begin{align}
\label{eqn:achcolsec} \ssum_{k=1}^K \Rsec_k &= \mmax{\CM_\Ks-\CW_\Ks}\\
\label{eqn:achcolx} \ssum_{k=1}^K \paren{\Rpub_k+\Rx_k} &= \CW_\Ks \\
\label{eqn:achcolM} \ssum_{k \in \Ss} \paren{\Rsec_k+\Rpub_k+\Rx_k} &\le \CM_\Ss,
	&&\forall \Ss \subseteq \Ks
\end{align}
so that we show the achievability of the boundary, which can be done by relabeling some of the open or extra messages as secret.  Clearly, lower secrecy rates are also thus achieved.  From \eqref{eqn:achcolM} and the GMAC coding theorem, with high probability the receiver can decode the codewords with low probability of error. Define $\XmS=\sum_{k=1}^K h_k \Xm_k$, and write
\begin{align}
H(\Wmsec_\Ks|\Zm) =H(\Wmsec_\Ks,\Zm)-H(\Zm) \hspace{-1.65in}&\\
	&=H(\Wmsec_\Ks,\XmS,\Zm)-H(\XmS|\Wmsec_\Ks,\Zm)-H(\Zm) \\
	&=H(\Wmsec_\Ks)+H(\Zm|\Wmsec_\Ks,\XmS)-H(\Zm) \notag \\
		&\hspace{.3in} +H(\XmS|\Wmsec_\Ks)-H(\XmS|\Wmsec_\Ks,\Zm) \\
	&= H(\Wmsec_\Ks)- I(\XmS;\Zm)-I(\XmS;\Zm|\Wmi_\Ks) \label{eqn:achprfcol1} 
\end{align}
where we used $\Markov{\Wmsec_\Ks}{\XmS}{\Zm} \Rightarrow H(\Zm|\Wmsec_\Ks,\XmS)=H(\Zm|\XmS)$ to get \eqref{eqn:achprfcol1}.  We will consider the two mutual information terms individually.  First, we have the trivial bound due to channel capacity: $I(\XmS;\Zm) \le n\CW_\Ks$.  We write the second out as $I(\XmS;\Zm|\Wmsec_\Ks) = H(\XmS|\Wmsec_\Ks)-H(\XmS|\Wmsec_\Ks,\Zm)$.  Since user $k$ sends one of $\Mpub_k \Mx_k$ codewords for each message, $H(\XmS|\Wmsec_\Ks) =n \sum_{k=1}^K \paren{\Rpub_k+\Rx_k} = n\CW_\Ks$ from \eqref{eqn:achcolx}.  We can also write $H(\XmS|\Wmsec_\Ks,\Zm) \le n\delta_n$ where $\delta_n \tozero$ as $n \toinf$ since, the eavesdropper can decode $\XmS$ given $\Wmsec_\Ks$ due to \eqref{eqn:achcolx} and code construction. Using these in \eqref{eqn:achprfcol1}, we get
\begin{equation}
\DeltaC_\Ks \ge 1-\frac{n\CW_\Ks-n\CW_\Ks+n\delta_n}{n\sum_{k=1}^K \Rsec_k} = 1-\e
\end{equation}
where $\e = \frac{\delta_n}{\sum_{k=1}^K \Rsec_k} \tozero$ as $n \toinf$.

The corollary simply follows from the definition of $\delta$-achievability as in the proof of Corollary \ref{cor:achindd}. \hfill \QED

\subsection{TDMA}
\label{app:achprfTDMA}
In the TDMA scheme described in Theorem \ref{thm:achTDMA}, only one user transmits at a time. Hence, $H(\Wsec_k|\Xm_{k^c},\Zm) = H(\Wsec_k|\Zm)$ as at any given time the codewords of the remaining users do not provide any information to the eavesdropper about the transmitting user's message.  As a result, both sets of secrecy constraints become equivalent.  Since this is a collection of single-user schemes, using the achievability proof in \cite{leung-hellman:gaussianwiretap}, and noting that the degradedness condition is only used for proving the converse, we can, for each user, achieve
\begin{align}
\Rsec_k &\le \mmax{\alpha_k g\paren{\frac{P_k}{\alpha_k}}-\alpha_k g\paren{\frac{h_k P_k}{\alpha_k}}} \\
\Rsec_k+\Rpub_k &\le \alpha_k g\paren{\frac{P_k}{\alpha_k}}
\end{align}
which, when simplified, gives \eqref{eqn:achTDMAP}.  We can use time-sharing between different scheduling schemes to achieve the convex closure of the union over all $\vg{\alpha}$ and power allocations.  The corollary follows from Definition \ref{def:deltaachrate}. \hfill \QED
\section{Outer Bounds}
\label{app:outprf}
We first adapt \cite[Lemma 10]{leung-hellman:gaussianwiretap} to upper bound the difference between the received signal entropies at the receiver and eavesdropper, when the eavesdropper's signal is degraded:

\begin{lemma}[Lemma 10 in \cite{leung-hellman:gaussianwiretap}]
\label{lem:outprfLeung}
Let $\xi = \ninv H(\Ym)$ where $\Ym,\Zm$ are as given in \eqref{eqn:YZdeg}. Then,
\begin{equation}
H(\Ym)-H(\Zm) \le n\xi - n\phi(\xi)
\end{equation}
where $\phi(\xi) \triangleq \onehalf\log\bracket{2\pi e\paren{1-h+\frac{h2^{2\xi}}{2\pi e}}}$.
\end{lemma}
\begin{corollary}
\label{cor:outprfLeung}
\begin{equation}
H(\Ym|\Xm_\Ss) - H(\Zm|\Xm_\Ss) \le n \paren{\CM_\Ssc-\CW_\Ssc}
\end{equation}
\end{corollary}
\begin{proof}
The proof is easily shown using the entropy power inequality, \cite{cover-thomas:IT}:
Recall that $H(\Zm)=H(\sqrt{h}\Ym+\NMW)$.  Then, by the entropy power inequality
\begin{equation}
2^{\frac{2}{n} H(\Zm)} = 2^{\frac{2}{n} H(\sqrt{h}\Ym+\NMW)} 
	\ge 2^{\frac{2}{n} [H(\Ym)+n\log\sqrt{h}]} +2^{\frac{2}{n} H(\NMW)}
\end{equation}
Now $H(\Ym) = n\xi$ and $H(\NMW) = \frac{n}{2} \log [ 2 \pi e(1-h)]$.  Hence,
\begin{equation}
2^{\frac{2}{n} H(\Zm)} \ge h2^{2\xi} + 2 \pi e (1-h)
\end{equation}
which, after taking the log, gives
\begin{equation}
H(\Zm) \ge \frac{n}{2} \log \bracket{2\pi e\paren{1-h+\frac{h2^{2\xi}}{2 \pi e}  }}
\end{equation}
subtracting from $H(\Ym)=n\xi$ completes the proof of the lemma.
To see the corollary, write
\begin{equation}
H(\Ym|\Xm_\Ss) \le \frac{n}{2} \log \paren{2 \pi e (1+P_{\Ss^c})}
\end{equation}
Let $H(\Ym|\Xm_\Ss) = n\xi$.  Then, $\xi \le \onehalf \log \paren{2 \pi e (1+P_{\Ss^c})}$, and since $\phi(\xi)$ is a non-increasing function of $\xi$, we get $\phi(\xi) \ge \phi\paren{\onehalf \log\paren{2\pi e(1+P_{\Ss^c})}}$. Since $\{\Xm_k\}$ are independent, we can use the lemma with $\Ym \rightarrow \Ym|\Xm_\Ss$ and $\Zm \rightarrow \Zm|\Xm_\Ss$,
\begin{align}
H(\Ym|\Xm_\Ss) - H(\Zm|\Xm_\Ss) \hspace{-1.3in} &\\
	&\le \frac{n}{2} \log \paren{2\pi e (1+P_{\Ssc})} -
		\frac{n}{2} \log \bracket{2 \pi e \paren{1+hP_{\Ssc}}}  \\
	&=n \paren{\CM_\Ssc-\CW_\Ssc}
\end{align}
\end{proof}

We also present the following lemma that is valid for the general (non-degraded) case:
\begin{lemma}
\label{lem:outprflim}
Let $\Ss \subseteq \Ks$.  Then,
\begin{equation}
\label{eqn:outprflim}
I(\Wsec_\Ss;\Ym|\Xm_\Ssc,\Zm) \le H(\Ym|\Xm_\Ssc)-H(\Zm|\Xm_\Ssc) +n\e_n
\end{equation}
\end{lemma}
\begin{proof} $I(\Wsec_\Ss;\Ym|\Xm_\Ssc,\Zm)$
\begin{align}
	&= H(\Ym|\Xm_\Ssc,\Zm)-H(\Ym|\Wsec_\Ss,\Xm_\Ssc,\Zm) \notag \\
	&\le H(\Ym|\Xm_\Ssc,\Zm)-H(\Ym|\Wsec_\Ss,\Xm_\Ks,\Zm) \\
	&=H(\Ym|\Xm_\Ssc,\Zm)-H(\Ym|\Xm_\Ks,\Zm) \\
	&=I(\Xm_\Ss;\Ym|\Xm_\Ssc,\Zm) \\
	&=I(\Xm_\Ss;\Ym|\Xm_\Ssc)+I(\Xm_\Ss;\Zm|\Xm_\Ssc,\Ym) \notag \\
		&\hspace{.3in} -I(\Xm_\Ss;\Zm|\Xm_\Ssc) \\
	&\le I(\Xm_\Ss;\Ym|\Xm_\Ssc)+H(\Xm_\Ss|\Xm_\Ssc,\Ym) \notag \\
		&\hspace{.3in} -I(\Xm_\Ss;\Zm|\Xm_\Ssc) \\
	&\le I(\Xm_\Ss;\Ym|\Xm_\Ssc)-I(\Xm_\Ss;\Zm|\Xm_\Ssc)+n\e_n \label{eqn:outprflim2}\\
	&=H(\Ym|\Xm_\Ssc)-H(\Ym|\Xm_\Ks) - H(\Zm|\Xm_\Ssc) \notag \\
		&\hspace{.3in} +H(\Zm|\Xm_\Ks) + n\e_n \\
	&=H(\Ym|\Xm_\Ssc)-H(\Zm|\Xm_\Ssc) +n\e_n
\end{align}
where $\e_n \tozero$ as $n \toinf$ and $\e \tozero$.  In \eqref{eqn:outprflim2} we used $H(\Xm_\Ss|\Xm_\Ssc,\Ym) \le H(\Wsec_\Ss|\Xm_\Ssc,\Ym) \le n\e_n$ where $\e_n \tozero$ as $n \toinf$ from Fano's inequality.  The last step comes from $H(\Ym|\Xm_\Ks)=H(\NM)=H(\Zm|\Xm_\Ks)=H(\NW)$.
\end{proof}

\subsection{Individual Constraints}
\label{app:outprfind}
The proof is a simple extension of the proof of Lemma 7 in \cite{leung-hellman:gaussianwiretap}, but stronger in the sense of \cite{maurer-wolf:weaktostrongsecrecy} as we prove an outer bound satisfying $H(\Wsec_k|\Xm_{k^c},\Zm) \ge H(\Wsec_k)-\e$, for all $k\in \Ks$.  Clearly, any set of rates satisfying the original individual constraints also satisfies these constraints.  We begin with:
\begin{lemma}
\label{lem:outprfind}
The individual secrecy rates must satisfy:
\begin{equation}
\label{eqn:outprfind1}
\Rsec_k \le \ninv \paren{H(\Ym|\Xm_{k^c})-H(\Zm|\Xm_{k^c})} +\e_n
\end{equation}
\end{lemma}
\begin{proof}
Note that $H(\Wm_\Ss|\Xm_{\Ss^c},\Ym,\Zm) \le H(\Wm_\Ss|\Ym) \le n \e'_n$
for all $\Ss \subseteq \Ks$ from Fano's inequality.  Using $H(\Wsec_k|\Xm_{k^c},\Zm) \ge H(\Wsec_k)-\e=n\Rsec_k-\e$, we can write
\begin{align}
n \Rsec_k &\le H(\Wsec_k|\Xm_{k^c},\Zm) +\e \\
	&\le H(\Wsec_k|\Xm_{k^c},\Zm) + n\e''_n - H(\Wsec_k|\Xm_{k^c},\Ym,\Zm) \\
	&= I(\Wsec_k;\Ym|\Xm_{k^c},\Zm) + n\e''_n
\end{align}
and we can then use Lemma \ref{lem:outprflim} with $\Ss=\{k\}$.
\end{proof}
When $\Zm$ is a degraded version of $\Ym$, Corollary \ref{cor:outprfLeung} gives
\begin{equation}
\Rsec_k \le \CM_k-\CW_k = g\paren{\frac{(1-h)P_k}{1+hP_k}}
\end{equation}

Corollary \ref{cor:outindd} follows by noting that a rate $\Rmd$ is achievable iff all $\Rd_k$ are such that $(\delta \Rd_k,\Rd_k)$ is achievable for user $k$. \hfill \QED

\subsection{Collective Constraints}
\label{app:outprfcol}
We show that any achievable rate vector, $\Rm$, needs to satisfy Theorem \ref{thm:outcol}.  We start with a lemma similar to Lemma \ref{lem:outprfind}:
\begin{lemma}
\label{lem:outprfcol}
\begin{equation}
\ssum_{k=1}^K \Rsec_k \le \ninv \paren{H(\Ym)-H(\Zm)} + \e_n
\end{equation}
where $\e_n \tozero$ as $\e \tozero$.
\end{lemma}
\begin{proof}
We have $H(\Wmsec_\Ks|\Ym,\Zm) \le H(\Wmsec_\Ks|\Ym) \le n \e'_n$ from Fano's inequality.  Using $H(\Wmsec_\Ks|\Zm) \ge H(\Wmsec_\Ks)-\e=n\sum_{k=1}^K \Rsec_k-\e$, we can write
\vspace{-11pt}
\begin{align}
n \ssum_{k=1}^K \Rsec_k &\le H(\Wmsec_\Ks|\Zm) +\e \\
	&\le H(\Wmsec_\Ks|\Zm) + n\e_n - H(\Wmsec_\Ks|\Ym,\Zm) \\
	&= I(\Wmsec_\Ks;\Ym|\Zm) + n\e_n
\end{align}
and we can then use Lemma \ref{lem:outprflim} with $\Ss=\Ks$.
\end{proof}
When $\Zm$ is a degraded version of $\Ym$, Corollary \ref{cor:outprfLeung} gives
\begin{equation}
\ssum_{k=1}^K \Rsec_k \le \CM_\Ks-\CW_\Ks = g\paren{\frac{(1-h)P_\Ks}{1+hP_\Ks}}
\end{equation}

\section{Sum Capacity for Degraded Case}
\label{app:degsumcap}
For the individual constraints, we can find the following limit on sum capacity by noting that $\DeltaI_k \ge 1$ for all $k \in \Ks$ implies $\DeltaI_\Ks \ge 1$ as shown in Section \ref{sec:defind}.  Hence, when the individual constraints are satisfied, the collective constraint must also be satisfied.  Thus, we can write
\begin{equation}
\sum_{k=1}^K \Rsec_k \le H(\Wsec_\Ks|\Zm) +\e \le I(\Wsec_\Ks;\Ym|\Zm)+\e
\end{equation}
as in the proof of Lemma \ref{lem:outprfcol} and use Lemma \ref{lem:outprflim} to get
$\sum_{k=1}^K \Rsec_k \le \CM_\Ks-\CW_\Ks$.

For the collective constraint, the outer bound was given in Theorem \ref{thm:outcol} as $\CM_\Ks-\CW_\Ks$.  The scheme to get the achievable rates given in \ref{thm:achcol} achieves this rate, and hence is the sum-capacity for collective constraints.

From Theorem \ref{thm:achTDMA}, we see that TDMA achieves a secrecy sum-rate of
$\sum_{k=1}^K \alpha_k g\paren{\frac{(1-h)P_k}{\alpha_k+hP_k}}$. 
Maximizing this over the time-sharing parameters $\{\alpha_k\}$, is a convex optimization problem over $\alpha_k$ whose solution is
\begin{equation}
\alpha_k^* = \frac{P_k}{\sum_{k=1}^K P_k}
\end{equation}
giving a sum rate of $\CM_\Ks-\CW_\Ks$.  Since the individual constraints are satisfied, this is an achievable sum-rate for the individual constraints, as well.  Hence, the sum-capacity for both sets of constraints is $\CM_\Ks-\CW_\Ks$. \hfill \QED
\section{Sum-Rate Maximization}
\newcommand{\rhodot}{\dot \rho}
\label{app:summaxprf}
\subsection{Optimum Powers}
\label{app:summaxprfopt}
We start with writing the Lagrangian to be minimized,
\begin{equation}
\label{eqn:summaxprfLag}
\Lag(\Pm,\muv) = \rho(\Pm)-\sum_{k=1}^2 \mu_{1k}P_k + \sum_{k=1}^2 \mu_{2k}(P_k-\Pmax_k)
\end{equation}
Equating the derivative of the Lagrangian to zero, we get
\begin{equation}
\label{eqn:summaxprfLagder}
\frac{\del \Lag(\Pmopt,\muv)}{\del \Popt_j} 	= \rhodot_j(\Pmopt) -\mu_{1j} + \mu_{2j} = 0
\end{equation}
where $\rhodot_j (\Pm) \triangleq \frac{h_j-\rho(\Pm)}{1+P_1+P_2}$.
It is easy to see that if $h_j > \rho(\Pmopt)$, then $\mu_{1j}>0$, and we have $\Popt_j=\Pmax_j$.  If $h_j < \rho(\Pmopt)$, then we similarly find that $\Popt_j=0$.  Finally, if $h_j = \rho(\Pmopt)$, we can have $0<\Popt_j<\Pmax_j$.  However, then $\rhodot_j(\Pmopt)=0$, so we can set $\Popt_j=0$ with no effect on the secrecy sum-rate.  Thus, we have $\Popt_j=\Pmax_j$ if $h_j < \rho(\Pmopt)$, and $\Popt_j=0$ if $h_j \ge \rho(\Pmopt)$. \hfill \QED

\subsection{Cooperative Jamming}
\vspace{-.05in}
\label{app:summaxprfcj}
\newcommand{\phidot}{\dot \phi}
The Lagrangian and its gradient are:
\vspace{-.05in}
\begin{equation}
\label{eqn:jamLag}
\Lag(\Pm,\muv) = \frac{\rho(\Pm)}{\phi_2(P_2)}
	-\sum_{k=1}^2 \mu_{1k}P_k + \sum_{k=1}^2 \mu_{2k}(P_k-\Pmax_k)
\end{equation}
\vspace{-.1in}
\begin{align}
\frac{\del \Lag}{\del \Popt_1} &= \label{eqn:jamLagder1}
\frac{\rhodot_1(\Pmopt)}{\phi_2(\Popt_2)}-\mu_{11}+\mu_{21} =0 \\
\frac{\del \Lag}{\del \Popt_2} &= \label{eqn:jamLagder2}
\frac{\rhodot_2(\Pmopt) \phi_2(\Popt_2)-\rho(\Pmopt)\phidot_2(\Popt_2)}
	{\phi_2^2(\Pmopt)} -\mu_{12} + \mu_{22} =0 \raisetag{.4in}
\end{align}
where $\phidot_2(P) \triangleq \frac{h_2-\phi_2(P)}{1+P}$.
Consider user 1.  The same argument as in the sum-rate maximization proof leads to $\Popt_1=\Pmax$ if $h_1 < \rho(\Pmopt)$ and $\Popt_1=0$ if $h_1 \ge \rho(\Pmopt)$.  Now we need to find $\Popt_2$.  We can write \eqref{eqn:jamLagder2} as
\begin{equation}
\frac{\del \Lag}{\del \Popt_2}=
\frac{\psi_2(\Pmopt)}{(1+\Popt_1+\Popt_2)^2 (1+h_2 \Popt_2)^2} -\mu_{12} +\mu_{22} =0
\end{equation}
where $\psi_2(\Pm) =P_1 h_2 (h_2-h_1) (P_2-p)(P_2-\bar p)$ and
\begin{gather}
\hspace{-.1in} p=\frac{-h_2(1-h_1) + \sqrt{D}}{h_2(h_2-h_1)}, \quad
\bar p =\frac{-h_2(1-h_1) - \sqrt{D}}{h_2(h_2-h_1)}\\
D=h_1h_2(h_2-1)\bracket{(h_2-1)+(h_2-h_1)P_1)}
\end{gather}

Note that the optimum power allocation for user 1 is equivalent to $\Popt_1=\Pmax$ if $h_1 < \phi_2(\Popt_2)$ and $\Popt_1=0$ if $h_1 \ge \phi_2 (\Popt_2)$.  Also observe that $\psi_2(\Pm)$ is an (upright) parabola in $P_2$.  
If $h_1<1$, we automatically have $\Popt_1=\Pmax$.  In addition, we have $\bar p<0$.  We first find when $\Popt_2=0$.  We see that $\psi_2(\Pm) \ge 0$ for all $\Pm \in \Ps$ if $p<0$, equivalent to having two negative roots, or $D<0 \Rightarrow h_2 \le \phi_1(\Pmax)$, equivalent to having no real roots of $\psi_2$.  Consider $0<\Popt_2<\Pmax$.  This is possible iff $\psi_2(\Pmopt)=0$.  Since $\Popt_1>0$, this happens only when $h_1=h_2$ or $\Popt_2=p>0$.  However, if $h_1=h_2$, we should be transmitting not jamming.  The last case to examine is when $\Popt_2=\Pmax$. This implies that $\psi_2(\v{\Pmax})<0$, and is satisfied when $p>\Pmax_2$.

Assume $h_2 \ge h_1 \ge 1$.  In this case, we are guaranteed $p\ge 0$.  If $\Popt_1=0$, then we must have $\Popt_2=0$ since the secrecy rate is 0.  If $h_1=h_2$, then regardless of $\Popt_2$, we have $\psi_2(\Pmax,\Popt_2)=0$ and jamming does not affect secrecy capacity, and we have $\Popt_2=0 \Rightarrow \Popt_1=0$.  Assume $h_2 > h_1$.  We would like to find when we can have $\Popt_1>0$. Since $h_1 < \phi_2(\Popt_2)$, we must have $\Popt_2>\frac{h_1-1}{h_2-h_1} \ge 0$, and $\psi_2(\Pmax,\Popt_2) \le 0$.  This implies $\bar p \le \Popt_2 \le p$.  It is easy to see that $\Popt_2=\min \braces{p,\Pmax}$ if $\frac{h_1-1}{h_2-h_1} < \min \braces{p,\Pmax_2}$ and $\Popt_2=0$ otherwise.

\bibliographystyle{IEEEtran}
\bibliography{IEEEabrv_mod,etekin_full}

\end{document}